\documentclass[12pt]{article}
\usepackage{graphicx}
\def\beq{\begin{equation}}
\def\eeq{\end{equation}}
\def\bea{\begin{eqnarray}}
\def\eea{\end{eqnarray}}
\def\nn{\nonumber}
\def\roughly#1{\mathrel{\raise.3ex\hbox
{$#1$\kern-.75em\lower1ex\hbox{$\sim$}}}}

\def\sss{\scriptscriptstyle}

\def\bd{B_d^0}

\def\bs{B_s^0}

\def\ks{K_{\sss S}}

\def\lft{{\sss L}}

\def\Aut{{\cal A}_{ut}}
\def\Act{{\cal A}_{ct}}
\def\Autp{{\cal A}'_{ut}}
\def\Actp{{{\cal A}'_{ct}}}

\def\aI{a_{\sss I}}
\def\aR{a_{\sss R}}

\def\btod{{\bar b} \to {\bar d}}
\def\btos{{\bar b} \to {\bar s}}
\def\plb#1#2#3{{ Phys.\ Lett.} {\bf #1B}, #3 (#2)}
\def\prd#1#2#3{{ Phys.\ Rev.} {\bf D#1}, #3 (#2)}

\def\prl#1#2#3{{ Phys.\ Rev.\ Lett.} {\bf #1}, #3 (#2)}

\begin{document}

\begin{flushright}  
\end{flushright}

\begin{center}
\bigskip
{\Large \bf \boldmath Getting SM and New Physics CP phases from $B$ 
Decays} \\
\bigskip
\bigskip
{\large Alakabha Datta \footnote{datta@physics.utoronto.ca}\\
\it Department of Physics, University of Toronto\\
\it 60 St.\ George Street, Toronto, ON, Canada M5S 1A7\\
}
\end{center}

\begin{center} 
\bigskip (\today)
\vskip0.5cm
{\Large Abstract\\}
\vskip3truemm
\parbox[t]{\textwidth} {A method to extract Standard Model(SM) CP 
phases involving B decays is presented.
The method involves a pair of decays where
 one decay  receives a significant $\btod$ penguin contribution
while the second has a significant $\btos$ penguin contribution,
but is dominated by a single amplitude. SM CP phase information is
obtained using the fact that the $\btos$ transition   amplitude is
related by SU(3) to a piece of the amplitude with the
$\btod$ transition. 
If there is significant new physics only in $\btos$ transitions 
and not in
$\btod$ transitions then this method can be used to measure new physics(NP)
phases.
}
\end{center}

\thispagestyle{empty}
\newpage
\setcounter{page}{1}
\baselineskip=14pt

\section{Introduction}

The $B$-factories BaBar and Belle have already reported a large number of
measurements involving $B$ decays, and this will continue for a number
of years. These measurements will test the SM picture of CP violation.
Deviations from the SM predictions will indicate new physics.
Hence, it becomes necessary to extract the SM phases  cleanly 
and in many different decays. If deviation from the SM and evidence for new 
CP violating phases are detected one would like to measure the new 
CP phases also.
In this talk we first present a method to measure SM phases which is 
clean and experimentally feasible. In the second part of the talk we discuss 
how our method can be use to measure NP phases if they are present.

The SM CP phase information is conveniently
represented by the  unitarity triangle,
whose interior angles are known as $\alpha$, $\beta$ and $\gamma$
\cite{pdg}. 
Many  methods
involving nonleptonic $B$ decays have been proposed for measuring the 
CP phases \cite{CPreview}. In general, these techniques suffer from some degree
of theoretical error due to the hadronic uncertainty 
in calculating nonleptonic $B$ decays.  Some
 methods are quite clean, i.e.\ they have little theoretical
uncertainty. An example is the extraction of $\beta$ in $\bd(t) \to J/\psi
\ks$ where the corrections from unknown hadronic quantities 
are highly CKM suppressed. 
{\footnote{$\bd(t) \to J/\psi
\ks$ measures only $\sin{2 \beta}$. The measurement of $\cos{2 \beta}$
is possible in $ B_d \to D^{(*)}D^{(*)} K$ decays \cite{BDDK}.
}}
 In general the extraction of CP angles require
a theoretical input about hadronic quantities \cite{LSS}.
To see this explicitly, consider processes where there is a $\btod$
penguin amplitude. Some examples of such processes are
$B_d \to \pi^+ \pi^{-}, B_d \to K_0 {\bar{K}}_0$ etc.

The general amplitude for the process $B^0 \to M_1 M_2$ involving, 
a $\btod$ penguin contribution, can be written
\bea
A(B^0 \to M_1 M_2) & = & A_u V_{ub}^* V_{ud}
 +  A_c V_{cb}^* V_{cd} \nn\\
&+& A_t V_{tb}^* V_{td} \
\eea
Using the CKM unitarity relation:
$$V_{ub}^* V_{ud} + V_{cb}^* V_{cd}+ V_{tb}^* V_{td} = 0,$$ one can 
 eliminate the $V_{tb}^* V_{td}$ term
to rewrite the amplitude as 
\bea
A& = & (A_u - A_t) V_{ub}^* V_{ud}  
+ (A_c - A_t) V_{cb}^* V_{cd} \nn\\
& \equiv & \Aut\ e^{i \gamma} e^{i \delta_{ut}} + \Act\ e^{i
  \delta_{ct}} \
\eea
The time-dependent measurement of $B^0(t)\to M_1 M_2$ allows one to
obtain the three observables
\bea
B &\equiv & \frac{1}{2} \left( |A|^2 + |{\overline A}|^2 \right) = 
\Act^2 + \Aut^2 \nn\\
& + & 2 \Act \, \Aut \cos\delta \cos\gamma ~ \nn \\
a_{dir} &\equiv & \frac{1}{2} \left( |A|^2 - |{\overline A}|^2 \right)
= - 2 \Act \, \Aut \sin\delta \sin\gamma ~\nn\\
\aI &\equiv & {\rm Im}\left( e^{-2i \beta} A^* {\overline A} \right) \nn\\
&- &\Act^2 \sin 2\beta 
 -  2 \Act \, \Aut \cos\delta \sin (2 \beta + \gamma)
\nn\\
&-& \Aut^2 \sin (2\beta + 2 \gamma)~ \
\eea
where ${\delta}\equiv {\delta}^{ut} - {\delta}^{ct}$

The three independent observables depend on five theoretical
parameters: $\Aut$, $\Act$, $\delta$, $\beta$, $\gamma$. Therefore one
cannot obtain CP phase information from these measurements.
Hence theoretical input about
hadronic quantities is necessary to get the CKM phase 
information. 

One popular class of methods relies on flavor SU(3) symmetry \cite{su3}
to provide the theoretical input to get CP phase information.
To see how this works, consider the decays
 $B_d \to \pi^+ \pi^-$ which has a $\btod$ penguin  and 
$B_s \to K^+ K^-$ which has a $\btos$ penguin\cite{Fl1}
 
The amplitudes for $B_d \to \pi^+ \pi^-$   and 
$B_s \to K^+ K^-$ can be written as
\bea
A_{\pi \pi} & = & 
 T^{\pi}\ e^{i \gamma} e^{i \delta_{ut}}
|V_{ub}^* V_{ud}|  + P^{\pi}|V_{cb}^* V_{cd}| \ e^{i
  \delta_{ct}} ~\nn\\
& & \nn\\
A_{KK} & = & 
 T^{K}|V_{ub}^* V_{us}| \ e^{i \gamma} e^{i \delta_{ut}'} + 
P^{K}|V_{cb}^* V_{cs}|\ e^{i
  \delta_{ct}'} ~\
\label{Bpipi}
\eea 
In these equations there 8 unknowns
which may be reduced to 7 by assuming $\beta$ from $B_d \to J/\psi K_s$ 
but there are only 6 measurements.
The theoretical input in the form of the
 SU(3) relation
$$ {P^{K} \over T^{K}}={P^{\pi} \over T^{\pi}} $$
allows us to extract $\gamma$.

This method has several potential  problems. First it requires
many measurements which can lead  to significant errors
in the extraction of the CP angle.
Moreover,
we want to estimate or reduce SU(3) breaking which is
around $\sim$ 25 \% and can also seriously affect the clean measurement of 
angles of the unitarity triangle.
Finally decays involving
 $B_s$ mesons may be experimentally difficult because of large
$B_s$ mixing and more importantly $B_s$ mesons are not
available at present B factories.

\section{New Method for measuring SM CP phase}

We now describe the new method to obtain SM phase information which is 
described in details in Ref.~\cite{Bpenguin} and builds on earlier works
presented in Ref.~\cite{BKKbar,BDDbar}.
This method uses
 SU(3) and neglects amplitudes that are highly CKM suppressed. This
reduces the number of measurements making it easier to obtain the
CKM unitarity angles from an experimental point of view.
With additional measurements one can reduce SU(3) breaking errors
allowing for more precise measurements of the angles of the unitarity triangle.
If annihilation type diagrams are neglected one can use many decays
that do not involve $B_s$ decays and hence would be  accessible at 
present B factories.

While, in principle the CKM suppressed amplitudes may become significant
if multiplied by highly enhanced strong amplitudes, this is unlikely
from our experience of hadron dynamics. More importantly the assumption 
that the highly CKM suppressed amplitudes may be neglected can be tested in 
experiments and
present experimental data appear to support this\cite{Bpenguin}.
Furthermore, whether annihilation type contributions are important 
can also be tested experimentally. In fact these contributions are typically
power suppressed $ \sim {\Lambda_{QCD} \over m_b}$ and are expected to be negligible specially for vector-vector final states. 

To see how this method works, consider the
the time-dependent measurement of $B^0(t)\to M_1 M_2$ involving
$\btod$ penguin. Experimentally we
obtain the three observables
\bea
B &\equiv & \frac{1}{2} \left( |A|^2 + |{\overline A}|^2 \right) = 
\Act^2 + \Aut^2 \nn\\
& + & 2 \Act \, \Aut \cos\delta \cos\gamma ~ \nn \\
a_{dir} &\equiv & \frac{1}{2} \left( |A|^2 - |{\overline A}|^2 \right)
= - 2 \Act \, \Aut \sin\delta \sin\gamma ~\nn\\
\aI &\equiv & {\rm Im}\left( e^{-2i \beta} A^* {\overline A} \right) \nn\\
&- &\Act^2 \sin 2\beta 
 -  2 \Act \, \Aut \cos\delta \sin (2 \beta + \gamma)
\nn\\
&-& \Aut^2 \sin (2\beta + 2 \gamma)~ \
\eea
However, one can partially solve the equations to obtain
\bea
\Act^2 = { \aR \cos(2\beta + 2\gamma) - \aI \sin(2\beta + 2\gamma) - B
\over \cos 2\gamma - 1} ~
\label{gammacond}
\eea
\bea
\aR^2 = B^2 - a_{dir}^2 - \aI^2 ~
\eea
 Hence if $\Act$ is known then we can find $\gamma$. We can obtain
 $\Act$ from a partner process related by SU(3).

 Consider now a decay $B' \to M'_1 M'_2$ involving a  $\btos$
penguin. 
We refer to this as the ``partner process.''

The general amplitude for $B' \to M'_1 M'_2$ can be written
\bea
A' & = & A'_u V_{ub}^* V_{us}
 +  A'_c V_{cb}^* V_{cs} 
+ A'_t V_{tb}^* V_{ts} \nn\\
& = & (A'_u - A'_t) V_{ub}^* V_{us}  
+ (A'_c - A'_t) V_{cb}^* V_{cs} \nn\\
& \equiv & \Autp\ e^{i \gamma} e^{i \delta'_{ut}} + \Actp\ e^{i
  \delta'_{ct}} \
\eea
We now assume $\Autp \ll
\Actp$ based on the fact that

$\left\vert {V_{ub}^* V_{us} / V_{cb}^*
V_{cs}} \right\vert \simeq 2\%$.

That is, {\it the partner process is assumed to be dominated
by a single amplitude.}
\bea
A' & \approx & \Actp\ e^{i \delta'_{ct}} 
\label{Bfampprime}
\eea
 Hence $\Actp$ is obtained just from the rate of $B' \to M'_1M'_2$ and no
  time dependent measurement is necessary thus greatly 
simplifying the experimental situation.

Now, in the SU(3) limit $\Act= \lambda\Actp$ 
where $\lambda=0.22$ is the Cabibbo angle and so we can get $\gamma$
from
\bea
{\Act}^2 &= &\lambda^2 {\Actp}^2\nn\\
&=& { \aR \cos(2\beta + 2\gamma) - \aI \sin(2\beta + 2\gamma) - B
\over \cos 2\gamma - 1} ~
\eea

There are many decays to which this method can be applied: the list of decays
where the final state can be pseudoscalar-pseudoscalar(PP),
pseudoscalar-vector(VP) and vector-vector(VV) are

\begin{enumerate}

\item $\bd \to D^+ D^-$ and $\bs \to D_s^+ D_s^-$, $\bd \to D_s^+
  D^-$, or $B_u^+ \to D_s^+ {\bar D}^0$;

\item $\bd \to \pi^+ \pi^-$ and $\bs \to K^0 {\bar K}^0$ or $B_u^+ \to
  K^0 \pi^+$;

\item $\bd \to K^0 {\bar K}^0$ and $\bs \to K^0 {\bar K}^0$ or $B_u^+
  \to K^0 \pi^+$;

\item $\bd\to\rho^0\rho^0$ and $\bd\to K^{*0}\rho^0$;

\item $\bs \to {\bar K}^{*0} \rho^0$ and $\bd \to K^{*0}\rho^0$;

\item $\bs \to \phi {\bar K}^{*0}$ and $\bd\to\phi K^{*0}$, $B_u^+ \to
\phi K^{*+}$, or $\bs\to\phi\phi$.

\end{enumerate}

The decays
\begin{enumerate}
\item $\bd \to D^+ D^-$ and , $\bd \to D_s^+
  D^-$, or $B_u^+ \to D_s^+ {\bar D}^0$;

\item $\bd \to \pi^+ \pi^-$ and  $B_u^+ \to
  K^0 \pi^+$;
\end{enumerate}
 are particularly interesting as data on these processes are
 available and can be/is being used already.

Finally we can address the issue of dropping the highly suppressed CKM 
contributions.
This dynamical assumption can be tested  in a variety of ways.
For instance if 
 $B' \to M'_1 M'_2$ is not dominated by a single amplitude and 
there is a significant subleading piece
then one would observe non zero Direct CP asymmetry 
or
measure non zero  T-violation asymmetry in VV modes \cite{TPpaper}.

One can use this method to considerebaly reduce
 SU(3)  by considering two pairs of
processes.
For example,
consider $\bd \to D^{+*} D^{-*}$ and  $\bd \to D_s^{+*}D^{-*}$ \cite{BDDbar}
where
we can consider pairs with different helicity states (i)
The theoretical input in this case is 

${\Act}_i = \lambda {\Actp}_i$ \quad 
${{\Act}_i \over  {\Act}_j}= {{\Actp}_i \over  {\Actp}_j}$

It can be shown that the leading SU(3) ( in $1/N_c$ approach) 
cancels in double ratio leaving a small residual SU(3) breaking. 

\section{ Measuring New Physics CP phases}
CP violation in the standard model is large and therefore one expects
large angles of the unitaritry triangle as well as significant direct CP 
violating effects in many decays \cite{datta-dcp}.
There  are many reason to believe that the Standard Model is not a complete theory as it leaves several puzzles unresolved, specially in the flavour sector.
Since CP is not a symmetry of the SM there is no reason to believe that
any new physics would be CP symmetric. One would therefore
expect deviations from SM CP predictions, specially in rare decays
where NP can compete with the SM contribution.  Rare decays
where the SM CP violation are tiny are very useful to look for 
new physics signals as any non zero CP violation would be a smoking gun signal for new physics.
There are several hints of possible deviations from the SM 
\cite{Datta,DattaLon}
and several interesting methods have been proposed to measure the parameters
of the underlying new physics \cite{DattaLon}. It appears so far that there
may be significant
new physics in $ \btos $ transitions while the $\btod$ 
transitions are relatively unaffected by new physics.

To demonstrate how our method can be turned around to measure NP phases, 
 we consider a specific model of NP.
One model which has received much attention  is $Z$-mediated (or $Z'$-mediated)
flavour-changing neutral currents (FCNC's) \cite{ZFCNC}.
Here one introduces an additional vector-singlet charge $-1/3$ quark
$h$ and allows it to mix with the ordinary down-type quarks $d$, $s$
and $b$. Since the weak isospin of the exotic quark is different from
that of the ordinary quarks, FCNC's involving the $Z$ are induced. The
$Zb{\bar s}$ FCNC coupling, which leads to the $\btos$ transitions, is
parametrized by the independent parameter $U_{sb}^{\sss Z}$:
\beq
{\cal L}^{\sss Z}_{\sss FCNC} = - {g \over 2 \cos\theta_{\sss W}} \,
U_{sb}^{\sss Z} \, {\bar s}_\lft \gamma^\mu b_\lft Z_\mu ~.
\eeq
Note that it is only the mixing between the left-handed components of
the ordinary and exotic quarks which is responsible for the FCNC.
Furthermore, the $Z$ decays to all $q{\bar q}$ pairs for
$q=u,d,s,c$. These are effectively new contributions to the
electroweak penguin operators of the SM.

The new-physics weak phase arises because $U_{sb}^{\sss Z}$ can be
complex -$U_{sb}$ can contain a CP violating phase $\Phi$.
However, because this parameter is universal, the weak phase
of all NP operators will be the same. 
Note that
$U_{db}$ is constrained from $B_d$ mixing so there is possible
 significant NP only 
in $ \btos$ decays.

Now  consider a $\btos $ processes, some examples of which are 
 $B_s \to K^0 \bar{K}^0$, $B_d \to K^{0*} \rho^0$ etc.
The general amplitude for such a process including the NP contribution is
\bea
A' = \Actp\ e^{i\delta'_{ct}} +
A_{NP} e^{i \Phi}\ e^{i \delta_{NP}}
\label{npa} 
\eea
Now as before there are three measurements: BR, $a_{direct}$ and $a_{indirect}$
and we assume
$B_s$ mixing is measured in $B_s \to J/\psi \eta(\phi)$.
There are four unknowns, $\Actp$, $A_{NP}$, $ \Phi$ and 
$\delta'=\delta'_{ct}-\delta_{NP}$, in Eq.~\ref{npa}. If we use
 $\Actp$ as input we can  solve for the rest including the NP phase
$\Phi$.

To obtain  $\Actp$, consider the  $\btod $  partner process: 
 $B_d \to K^0 \bar{K}^0$, $B_d \to \rho^0 \rho^0$ etc.
The amplitude here is 
\bea
A& = &  \Aut\ e^{i \gamma} e^{i \delta_{ut}} + \Act\ e^{i
  \delta_{ct}} \
\label{sma}
\eea
Now $\beta$ is obtained from
$\bd(t) \to J/\psi
\ks$ and $\gamma$ may be obtained from $ B \to \pi \pi$ \cite{GL}
as it is a $\btod$ transition and is not significantly affected by new
physics according to our assumption. The three unknowns
$\Act$, $\Aut$ and $\delta=\delta_{ct}-\delta_{ut}$ can then be obtained  
from the 3 time independent measurements. One can the use SU(3) symmetry to obtain
$\Actp$ in Eq.~\ref{npa} from
$\Act= \lambda \Actp$.

 Note that this model of NP  gives the same $\Phi$ 
from pairs $B_s \to K^0 \bar{K}^0$ and $B_d \to K^0 \bar{K}^0$ 
and $B_s \to K^{0*} \rho^0$ and $B_d \to \rho^0 \rho^0$ and so on.
Hence if this is not found to be the case this model of NP
would be ruled out.
 
There are models of NP which contain more then one NP operator.
The general treatment of this case is described in Ref.~\cite{DattaLon}.
As an example of NP model with more than one NP operator, consider
supersymmetric models with broken R-parity.
The $L$-violating couplings in this model are 
given by \cite{DatXin}
\bea
{\cal L}_{\lambda^{\prime}}&=&-\lambda^{\prime}_{ijk}
\left [\tilde \nu^i_L\bar d^k_R d^j_L+\tilde d^j_L\bar d^k_R\nu^i_L
       +(\tilde d^k_R)^*(\bar \nu^i_L)^c d^j_L\right.\nonumber\\
& &\hspace{1.5cm} \left. -\tilde e^i_L\bar d^k_R u^j_L
       -\tilde u^j_L\bar d^k_R e^i_L
       -(\tilde d^k_R)^*(\bar e^i_L)^c u^j_L\right ]+h.c.\
\label{Lviolating}
\eea

{}From this Lagrangian, we see that there are $R$-parity-violating
contributions to all $\btos$ transitions \cite{lambdabNP}. There is a
single contribution to the decay ${\bar b}\to {\bar s} u {\bar{u}}$:
\beq
L_{eff}= -\frac{\lambda^{\prime}_{i12} \lambda^{\prime*}_{i13}} {2 m_{
\widetilde{e}_i}^2} \bar u_\alpha \gamma_\mu \gamma_L u_\beta \, \bar
s_\beta \gamma_\mu \gamma_R b_\alpha ~.
\eeq
For $\btos\ d{\bar{d}}$, there are four terms:
\bea
L_{eff}&=&
\frac{\lambda^{\prime}_{i11} \lambda^{\prime*}_{i23}}{  m_{ \widetilde{\nu}_i}^2}
\bar d \gamma_L d_\beta \,
\bar {s}\gamma_R b
+\frac{\lambda^{\prime}_{i32} 
\lambda^{\prime*}_{i11}}{  m_{ \widetilde{\nu}_i}^2}
\bar d \gamma_R d \, \bar s \gamma_L b \nonumber \\
&-&\frac{\lambda^{\prime}_{i12} \lambda^{\prime*}_{i13}}
{2  m_{ \widetilde{\nu}_i}^2}
\bar d_\alpha \gamma_\mu \gamma_L d_\beta \, \bar s_\beta \gamma_\mu 
\gamma_R b_\alpha
-\frac{\lambda^{\prime}_{i31} \lambda^{\prime *}_{i21}}
{2  m_{ \widetilde{\nu}_i}^2}
\bar d_\alpha \gamma_\mu \gamma_R d_\beta \, \bar s_\beta \gamma_\mu \gamma_L 
b_\alpha ~.
\label{Lbsdd}
\eea
Finally, the relevant Lagrangian for the $\btos\ s {\bar{s}}$\cite{Datta}
transition is
\beq
L_{eff} = \frac{\lambda^{\prime}_{i32} \lambda^{\prime*}_{i22}} { m_{
\widetilde{\nu}_i}^2} \bar s \gamma_R s \, \bar {s}\gamma_L b+
\frac{\lambda^{\prime}_{i22} \lambda^{\prime*}_{i23}} { m_{
\widetilde{\nu}_i}^2} \bar s \gamma_L s \, \bar {s}\gamma_R b ~.
\eeq

{}From the above expressions we can deduce the following predictions
of $R$-parity-violating SUSY models. First, since there is only a
single term contributing to $\btos\ u {\bar u}$ transitions, the
measured value of the new CP phase, $\Phi_{uu}$,
 should be independent of the decay pairs
considered with the same underlying
$\btos\ u {\bar u}$ transition.
On the other hand, since there is more than one contribution
to both $\btos\ d {\bar d}$ and $\btos\ s {\bar s}$,  the  the value of
effective CP phases $\Phi_{dd}$ and $\Phi_{ss}$ \cite{DattaLon}
 will be process-dependent. Should this
pattern of NP weak phases not be found experimentally, we can rule out
this model of new physics.

\section{Conclusions}
In conclusion we have
have presented a new method based on SU(3) and a testable dynamical 
assumption to extract the angles of
 of the the unitarity triangle.
This method can  be applied to many pairs of decays, many of which do 
not involve $B_s$ decays and can be used in present B-factories.
This method can be used to considerably reduce SU(3) breaking error
and finally the method can be turned around to measure new physics parameters
and rule out or constrain many models of new physics.

{\bf Acknowledgements}:
This work was financially supported by NSERC of Canada.


\end{document}